\documentclass[twocolumn]{jpsj2}

\newcommand{\p}[1]{\psi_{#1}}
\newcommand{\pd}[1]{\psi_{#1}^\dagger}
\newcommand{\pp}[1]{\Psi_{#1}}
\newcommand{\ppd}[1]{\Psi_{#1}^\ast}
\newcommand{\ket}[1]{|{#1}\rangle}
\newcommand{\aho}{a_\mathrm{HO}}
\newcommand{\oho}{\omega_\mathrm{HO}}

\newcommand{\ee}{\mathrm{e}}
\newcommand{\ii}{\mathrm{i}}
\newcommand{\rr}{\mathbf{r}}

\newcommand{\U}{\mathcal{U}}
\newcommand{\Ud}{\mathcal{U}^\dagger}
\newcommand{\D}{\mathcal{D}}
\newcommand{\ti}[1]{\tilde{#1}}

\def\runtitle{Stability of Magnetically Trapped Bose-Einstein Condensates}
\def\runauthor{Yuki \textsc{Kawaguchi} and Tetsuo \textsc{Ohmi}}

\title{%
Stability of Magnetically Trapped Bose-Einstein Condensates
}

\author{%
Yuki \textsc{Kawaguchi}\thanks{E-mail: yuki@scphys.kyoto-u.ac.jp} and
Tetsuo \textsc{Ohmi}
}

\inst{%
Department of Physics, Graduate School of Science, Kyoto University, Kyoto 606-8502
}

\recdate{\today}

\abst{%
According to the adiabatic approximation atoms moving in a magnetic trap keep their magnetic states.
We investigate the validity of this approximation for quantum condensates,
where a change of field's direction generates effective interactions between hyperfine angular momentum states.
Condensates in general traps are found to be stable
because they are confined in the vicinity of the trap center.
A decay of a condensate is observable in a trap with extremely large field gradient.

}

\kword{%
Bose-Einstein condensation, magnetic trap, Gross-Pitaevskii Equation, spin-flip, adiabatic approximation
}

\begin{document}
\sloppy
\maketitle

\section{Introduction}

Bose-Einstein condensate (BEC) of ultracold atoms has been studied furiously
since its experimental achievements~\cite{Anderson1995,Davis1995,Bradley1997}.
It still provides us a lot of interesting features.
Recently a quite small condensate on a microelectronic chip was realized,
where modest electric currents produced large magnetic field gradients
and a microscopic trap was formed on a chip~\cite{Hansel2001}.
A quantized vortex in a condensate is also one of the interesting phenomena.
A method of creating a vortex topologically through inverting the bias field along the trap axis
was suggested~\cite{Isoshima2000} and experimentally confirmed~\cite{Leanhardt2002}.
Since the field gradients used in these experiments are large as compared to the bias fields,
spin of a condensed atom may flip in the trap.

According to the adiabatic approximation,
an ultracold atom moving in a spatially changing magnetic field
keeps the relative direction of its magnetic moment to the field.
Atoms are classified into three states by the sign of Zeeman energy:
weak-field-seeking state~(\emph{wfss}),
neutral state~(\emph{ns}) and strong-field-seeking state~(\emph{sfss}).
In a magnetic field which has a minimum point of the magnitude in space,
only the atoms in \emph{wfss} can be trapped.

The adiabatic approximation requires that the field's direction $\theta$
which trapped atom feels
changes slower than the precession of the magnetic moment:
\begin{eqnarray}
 \frac{d\theta}{dt} < \omega_{\rm Larmor}.
\end{eqnarray}
This adiabatic condition is violated in regions of very small magnetic fields.
The motion of spin in a trap with no bias field has been calculated classically~\cite{Antillon1992}.
It was found that the size of such non-adiabatic region is very small
and the trapped atoms satisfies the adiabatic condition.

In this paper we consider quantum condensates.
When we describe the Gross-Pitaevskii (GP) equation with 
order parameters corresponding to \emph{wfss}, \emph{ns} and \emph{sfss}~\cite{Ohmi1998},
coupling terms between these states appear in the equation, besides nonlinear terms.
The adiabatic approximation requires these coupling terms to be negligible.
We evaluate the effects of these coupling and
investigate the limit of the application of the adiabatic approximation
by calculating the lifetime of a condensate.

We treat alkali atoms with a hyperfine state $|{\mathbf F}|=1$ such as $^{23}$Na and $^{87}$Rb.
In the next section we briefly introduce the treatment of BEC with internal degrees of freedom.
we represent the GP equation for \emph{wfss}, \emph{ns} and \emph{sfss},
and derive the expression of the decay rate.
In Sec.~\ref{sec:discuss} the lifetime of a condensate is calculated numerically.
We discuss about its dependence on field geometries.
The effect of singularity of the condensate and population of the trapped atoms are also referred to.
The final section is devoted to conclusions.

\section{Bose-Einstein Condensation in a Magnetic Field}
\subsection{Gross-Pitaevskii equation of $|\mathbf{F}|=1$ BEC}

Let $\mathbf{F}$ be the angular momentum operator with hyperfine state $|\mathbf{F}|=1$.
The eigenvalues of $F_z$ are $1, 0$ and $-1$.
The corresponding eigenstates are $\ket{1}, \ket{0}$ and $\ket{-1}$,
which satisfy $F_z\ket{m}=m\ket{m}$.
In this basis, the order parameter $\ket{\Psi}$ is expanded as
\begin{eqnarray}
\ket{\Psi}=\sum_{m=0,\pm1}\pp{m}\ket{m}.
\end{eqnarray}
It is convenient in a certain case to take another basis set of $\ket{x}, \ket{y}$ and $\ket{z}$,
which satisfy $F_\alpha\ket{\alpha}=0\ (\alpha=x,y,z)$.
$\ket{\Psi}$ is represented in this basis as
\begin{eqnarray}
\ket{\Psi}=\sum_{\alpha=x,y,z}\pp{\alpha}\ket{\alpha}.
\end{eqnarray}
Clearly, the state $\ket{z}$ corresponds to $\ket{0}$
and other states satisfy the relations, given by
\begin{eqnarray}
\ket{\pm1} = \mp \frac{1}{\sqrt{2}} \left( \ket{x}\pm \ii\ket{y} \right).
\end{eqnarray}

The Hamiltonian, which should be invariant
under spin space rotation and gauge transformation except for Zeeman term in the $xyz$-representation,
is described as
\begin{equation}
 \begin{split}
  H=&\int d^3r \left[\pd{\alpha}\left(-\frac{\hbar^2}{2M}\nabla^2-\mu\right)\p{\alpha}
	+\frac{g_1}{2}\left(\pd{\alpha}\p{\alpha}\right)^2 \right.\\
	&\left.+\frac{g_2}{2}\left|\p{\alpha}\p{\alpha}\right|^2
	-\ii\epsilon_{\alpha\beta\lambda}\gamma_\mu B_\lambda\p{\alpha}\pd{\beta} \right],
 \end{split}
\end{equation}
where $\gamma_\mu\simeq -\mu_B/2$ is the gyromagnetic ratio and
$\pd{\alpha},\p{\alpha}$ are the field operators for a Bose particle in the $\ket{\alpha}$ state.
The coupling constants are given by
\begin{eqnarray}
g_1=\frac{4\pi \hbar^2}{M}a_2, \hspace{5mm} g_2=\frac{4\pi\hbar^2}{M}\frac{a_0-a_2}{3}.
\end{eqnarray}
The scattering lengths for $^{87}$Rb are $a_0=5.82$~nm and $a_2=5.66$~nm~\cite{Ho1998}.
From the equation of motion for $\p{\alpha}$, one obtains the Gross-Pitaevskii equation:
\begin{equation}
 \begin{split}
  \ii\hbar\frac{\partial \pp{\alpha}}{\partial t}
	=& \left(-\frac{\hbar^2}{2M}\nabla^2-\mu\right)\pp{\alpha}
	+g_1(\ppd{\beta}\pp{\beta})\pp{\alpha}\\
	&+g_2(\pp{\beta}\pp{\beta})\ppd{\alpha}+\ii\epsilon_{\alpha\beta\lambda}\gamma_\mu B_\lambda\pp{\beta},
\label{GP-xyz}
 \end{split}
\end{equation}
where the ground state average $\langle \p{\alpha} \rangle$ is replaced by $\pp{\alpha}$ 
and approximations such as $\langle \pd{}\p{}\p{} \rangle \sim \ppd{}\pp{}\pp{}$ are introduced.
The GP equation for $\pp{+1,0,-1}$ is obtained
by transforming $\pp{x,y,z}$ to $\pp{+1,0,-1}$ in eq.~\eqref{GP-xyz}.
When the field is homogeneous in space and parallel to $z$-axis,
the Zeeman term yields no coupling in the representation of $\pp{+1,0,-1}$
and is simply described as $|\gamma_\mu| B m\pp{m}$.

\subsection{Magnetically trapped BEC}

Let us consider a condensate in a magnetic trap.
The magnetic field must have a minimum point of its magnitude to trap atoms.
In most instances the magnetic traps used in laboratories, for example a Ioffe-Prichard trap,
make spatial change of the field's direction as well as the magnitude.
This change causes spin-flips.

We assume a Ioffe-Prichard trap with a quadrapole magnetic field
\begin{eqnarray}
\mathbf{B}_{\perp}(\rr) =
\begin{pmatrix}
B_{\perp}(r)\cos(-\phi) \\ B_{\perp}(r)\sin(-\phi) \\ 0
\end{pmatrix},
\end{eqnarray}
where $\rr=(r, \phi, z)$ is the cylindrical coordinates.
The magnitude of the quadrapole field is proportional to $r$ near the trap center:
$B_\perp(r) \sim B'r$ at $r \sim 0$.
Assuming that the trap size along to $z$-axis is much larger than its perpendicular size,
we approximate the $z$-component of the magnetic field to be uniform: $\mathbf{B}_0=B_0\hat{z}$.
This bias field $B_0$ should be strong enough
to satisfy the condition $|\gamma_\mu| B_0 \gg k_\mathrm{B}T$
so that the thermal fluctuation does not dominate in spin-flip processes.

In such a spatially changing magnetic field,
we chose the direction of the local magnetic field as a quantum axis
\begin{eqnarray}
\hat{l}=
(\sin\beta\cos\phi, \ -\sin\beta\sin\phi, \ \cos\beta).
\end{eqnarray}
Here $\beta$ is an angle between $z$-axis and the magnetic field, given by
\begin{eqnarray}
\beta = \arctan\left(\frac{B_\perp(r)}{B_0}\right) = \arctan\left(\frac{r}{R}\right),
\end{eqnarray}
where $R=B_0/B'$ is a characteristic length of the field gradient.
The state $\ket{w(\rr)}$ corresponding to \emph{wfss}, 
in which a hyperfine spin keeps its direction parallel to the magnetic field, satisfies
\begin{eqnarray}
(\mathbf{F}\cdot\hat{l})\ket{w(\rr)} = \ket{w(\rr)}.
\end{eqnarray}
Similarly, the states $\ket{n(\rr)}$ and $\ket{s(\rr)}$ representing \emph{ns} and \emph{sfss} satisfy
\begin{align}
(\mathbf{F}\cdot\hat{l})\ket{n(\rr)} &= 0, \\
(\mathbf{F}\cdot\hat{l})\ket{s(\rr)} &= -\ket{s(\rr)}.
\end{align}
These states $\ket{i(\rr)} (i=w,n,s)$ are written with $\ket{\alpha}$ as
\begin{align}
\ket{i(\rr)}=\sum_{\alpha=x,y,z}\U'_{i\alpha}\ket{\alpha}.
\label{ket-trans}
\end{align}
By the analogy with the relation between $\ket{1,0,-1}$ and $\ket{x,y,z}$,
the transform matrix $\U'$ is defined with the field's direction $\hat{l}$
(and therefore depends on space) as
\begin{eqnarray}
\U'=\ee^{-\ii\phi}
\begin{pmatrix}
-\frac{1}{\sqrt{2}}(\hat{m}+\ii\hat{n}) \\ \hat{l} \\ \frac{1}{\sqrt{2}}(\hat{m}-\ii\hat{n})
\end{pmatrix},
\end{eqnarray}
where $\hat{m},\hat{n}$ are real vectors satisfying $\hat{m}\times\hat{n}=\hat{l}$
and $(\hat{l},\hat{m},\hat{n})$ forms a triad.
The transformation has an arbitrary phase,
and we choose it as $-\phi$ so that the state $\ket{w(\rr)}$ has no singularity at the trap center ($\beta=0$).
Then the winding numbers of the states $\ket{n(\rr)}$ and $\ket{s(\rr)}$ are $-1$ and $-2$, respectively.

The order parameter $\ket{\Psi}$ can be expanded
with the states $\ket{i(\rr)}$ as
\begin{eqnarray}
\ket{\Psi}=\sum_{i=w,n,s}\pp{i}(\rr)\ket{i(\rr)},
\end{eqnarray}
The transformation between $\pp{\alpha}$ and $\pp{i}$ is given by
\begin{eqnarray}
\pp{\alpha}=(\U'_{i\alpha})^\mathrm{T}\pp{i} \equiv \U_{\alpha i}\pp{i},
\label{psi-trans}
\end{eqnarray}
where the subscripts $\alpha, i$ denote $x,y,z$ and $w,n,s$, respectively,
and repeated index is summed.
Hereafter we use these characters in this notation.

Substituting the transformation eq.\eqref{psi-trans} into eq.~\eqref{GP-xyz},
we can construct the GP equation for $\pp{i}$ as
\begin{equation}
\begin{split}
\ii\hbar\frac{\partial}{\partial t}\pp{i}
 =& -\frac{\hbar^2}{2M} \Ud_{i\alpha} \nabla^2(\U_{\alpha j}\pp{j}) -\mu\pp{i}\\
 & + g_1\left(\ppd{j}\pp{j}\right)\pp{i}\\
 & + g_2\left[\pp{k}(\U^\mathrm{T}\U)_{kl}\pp{l}\right](\U^\mathrm{T}\U)^\ast_{ij}\ppd{j} \\
 & - |\gamma_\mu| B(r) ( \ii \epsilon_{\alpha\beta\lambda}\hat{l}_{\lambda}\Ud_{i\alpha}\U_{\beta j} )\pp{j}.
\label{GP-wns}
\end{split}
\end{equation}
The first term in the right hand side describes the kinetic energy,
which yields the additional terms $\Ud(\nabla\U)\cdot(\nabla\Psi)+\Ud(\nabla^2\U)\Psi$.
These extra terms come from the spatial change of the field's direction
and cause coupling.
Since we have chosen the quantum axis in the direction of the magnetic field,
the Zeeman term does not yields coupling term
and the matrix elements are simplified as
\begin{eqnarray}
-\ii \epsilon_{\alpha\beta\lambda}\hat{l}_{\lambda}\Ud_{i\alpha}\U_{\beta j}
 = \delta_{ij}(\delta_{wi}-\delta_{si}).
\end{eqnarray}
The matrix $\U^\mathrm{T}\U$ also have a simple form
\begin{eqnarray}
\U^\mathrm{T}\U=\ee^{-2\ii\phi}
\begin{pmatrix} 0&0&-1 \\ 0&1&0 \\ -1&0&0 \end{pmatrix}.
\end{eqnarray}

The symmetry of the field suggests that $\pp{i}$ is independent of $z$.
Assuming a cylindrical symmetry of the condensate,
we divide $r$ and $\phi$-dependence of the order parameter as
\begin{eqnarray}
\pp{i}(\rr)=f_i(r,t)\ee^{\ii m_i\phi}.
\end{eqnarray}
From eq.\eqref{GP-wns} $\phi$-dependence of each order parameter should satisfy the relation $m_w=m_n=m_s\equiv m$.
This means that \emph{wfss} with winding number $m$ can couple
only with \emph{ns} with singularity $m-1$ and \emph{sfss} with singularity $m-2$.

When the distribution size of the trapped atoms is much smaller than $R$,
the Zeeman term is approximated as a harmonic potential
\begin{align}
\begin{split}
|\gamma_\mu| B(r) &= |\gamma_\mu| \sqrt{B_\perp(r)^2+B_0^2}\\
	&\simeq |\gamma_\mu| B_0 \left\{ 1 + \frac{1}{2}\left(\frac{r}{R}\right)^2 \right\}.
\label{zeeman}
\end{split}
\end{align}
It is convenient to scale lengths and energies
by the harmonic oscillator length $\aho$ and the energy $\hbar\oho$, where
\begin{align}
\aho&=\sqrt{\frac{\hbar}{M\oho}},\\
\oho&=B'\sqrt{\frac{|\gamma_\mu|}{MB_0}}.
\end{align}
The length $\aho$ is in the order of the distribution size of the condensate.
When we use $r$ and $f$ as dimensionless variables (scaling $r/\aho\to r, f\aho^{3/2}\to f$),
the GP equation is rewritten in a simple form, given by
\begin{equation}
\begin{split}
\frac{\ii}{\oho}\frac{\partial f_i}{\partial t}
=& -\frac{1}{2}\D_{ij}f_j - \ti{\mu}f_i + \ti{g}_1 (f_j^\ast f_j)f_i \\
&+ \ti{g}_2(f_n^2-2f_w f_s)\ee^{-2\ii\phi}(\U^\mathrm{T}\U)^\ast_{ij}f_j^\ast \\
&+ \ti{B}(r)(\delta_{wi}-\delta_{si})f_i,
\label{GP-dimensionless}
\end{split}
\end{equation}
where $\displaystyle{\ti{B}=|\gamma_\mu| B/\hbar\oho}$, 
$\displaystyle{\ti{\mu}=\mu/\hbar\oho}$, $\displaystyle{\ti{g_i}=g_i/\aho^3\hbar\oho}$.
The matrix $\D$ comes from the kinetic energy term $\nabla^2 (\U\pp{})$
and consists of the functions of $r,\phi$ and the differential operators with $r$.

\subsection{Lifetime of weak-field-seeking state}

To calculate the lifetime of trapped condensate, we consider the situation that
the only atoms in \emph{wfss} exist in the trap at the time $t=0$
and populations of \emph{ns} and \emph{sfss} increase with time.

Assuming $|f_w|\gg|f_n|,|f_s|$,
\emph{wfss} is regarded as a stationary state.
When we trap $N_0$ atoms par $\aho$ along $z$-axis and normalize $f_w(r)$ by
\begin{eqnarray}
 2 \pi \int_0^\infty rdr |f_w(r)|^2 =1,
\end{eqnarray}
\emph{wfss} obeys the equation
\begin{eqnarray}
\ti{\mu} f_w=-\frac{1}{2}\D_{ww}f_w+\ti{g}_1 N_0 {f_w}^3+\ti{B}(r)f_w,
\label{eq-fw}
\end{eqnarray}
where $\D_{ww}$ is given by
\begin{align}
\begin{split}
\D_{ww}=&\D_0(m) + \frac{2m(1-\cos\beta)}{r^2} \\
 &- \frac{\beta '^2}{2} + \frac{-7+8\cos\beta-\cos{2\beta}}{4r^2},
\end{split}\\
\D_0(m)\equiv&\frac{d^2}{dr^2}+\frac{1}{r}\frac{d}{dr}-\frac{m^2}{r^2}.
\end{align}
Similarly, the equation for \emph{ns} is also simplified as
\begin{align}
\left(\ti{\mu} + \frac{\ii}{\oho}\frac{\partial}{\partial t}\right)&f_n
=-\frac{1}{2}\left[\D_0(m-1)f_n+\D_{nw}f_w\right],
\label{ns}\\
\begin{split}
\D_{nw} &= \sqrt{2}\beta '\frac{d}{dr} + \sqrt{2}\frac{m\sin\beta}{r^2}\\
& + \frac{(\cos\beta-2)\sin\beta+r\beta '+r^2\beta ''}{\sqrt{2}r^2}.
\end{split}
\end{align}
Since the kinetic energy term $\D_{nn}$ has no singularity at $r=0$ except for $\D_0(m-1)$,
we neglect the additional terms in $\D_{nn}$.
We also neglect spin-flips to \emph{sfss} because
\emph{wfss} and \emph{sfss} do not interact at $r=0$.

It is very useful to expand $f_n$ with the $(m-1)$th Bessel function $J_{m-1}(kr)$ as
\begin{eqnarray}
f_n(r,t) = \int^\infty_{-\infty}d\omega \ee^{-\ii\omega t} \int^\infty_0 kdk J_{m-1}(kr)j_n(k,\omega).
\end{eqnarray}
Substituting this form into eq.~\eqref{ns} and
using the Bessel function's property:
\begin{eqnarray}
g(x)=\int^\infty_0 ydy J_m(yx)\int^\infty_0 zdz J_m(yz) g(z),
\end{eqnarray}
we obtain a description of $f_n(r,t)$, which should satisfy the initial condition $f_n(r,0)=0$, as
\begin{equation}
\begin{split}
&f_n(r,t)\\
&=\int^\infty_0 kdk J_{m-1}(kr) F_w(k)\frac{\exp[\ii(\ti{\mu}-k^2/2)\oho t]-1}{\ti{\mu}-k^2/2},
\end{split}
\end{equation}
where
\begin{align}
F_w(k)\equiv \frac{1}{2}\int^\infty_0 rdr J_{m-1}(kr) \D_{nw} f_w(r).
\label{Fw_k-def}
\end{align}
Then the lifetime of the condensate, or the rate of atoms which flip to \emph{ns} par unit time, is given by
\begin{align}
\begin{split}
\frac{1}{\tau}&\equiv \frac{2\pi}{t} \int^\infty_0 rdr |f_n(r,t)|^2 \\
 &= 4\pi^2\oho\left|F_w(\sqrt{2\ti{\mu}})\right|^2,
\end{split}
\end{align}
where the concerning time scale is much longer than the period of Larmor precession
and the approximation
$(\sin \omega t/\omega)^2\sim \pi t\delta(\omega)$ is applied.

The function $F_w(k)$ depends on the distributions of \emph{wfss} and \emph{ns},
and the lifetime $\tau$ does too.
Since the atoms in \emph{ns} do not feel Zeeman potential,
they are in an eigen-momentum state $J_{m-1}(kr)$ with energy $\ti{\mu}$,
which corresponds to the chemical potential of \emph{wfss}.
We show in the next section that
this energy $\ti{\mu}$ strongly affects the coupling.

\section{Calculation and Discussion}
\label{sec:discuss}
Here we introduce a new parameter $b\equiv R/\aho$,
which indicates the changing rate of the field in the region of \emph{wfss}'s distribution.
From the definition of $R$ and $\aho$, $b^2$ corresponds to the ratio of the Zeeman energy at the trap center
to the harmonic oscillator energy.
We investigate two cases of $b\gg 1$ and $b\simeq 1$, respectively.
In the most cases of experiments $b$ is much larger than 1,
for instance
$b=51$ for $^{87}$Rb in a trap of $B_0=1$~G, $B'=300$~G/cm
and $b=10$ in $B_0=1$~G, $B'=8$~kG/cm.

\subsection{Distribution in the small field gradient $b\gg 1$}
When the distribution of the condensate is restricted within a narrow region at the trap center, $b\gg 1$,
the approximation eq.~\eqref{zeeman} is valid.
In this limit the magnetic field is almost parallel to $z$-axis ($\beta\ll 1$).
The operator $\D_{ww}$ and $\D_{nw}$ can also be expanded with $1/b$.
Neglecting the higher order of $1/b$, these terms are simplified as
\begin{align}
\D_{ww}&=\D_0(m),\\
\D_{nw}&=\frac{\sqrt{2}}{b}\left(\frac{d}{dr}+\frac{m}{r}\right).
\end{align}
In this case, $F_w(k)$ is described with a simple form:
\begin{eqnarray}
F_w(k)=\frac{k}{\sqrt{2}b}\int^\infty_0rdr J_m(kr) f_w(r).
\label{Fw_k}
\end{eqnarray}
It should be noted that using the symmetry of the Bessel function $J_{-m}(kr)=(-1)^mJ_m(kr)$,
the lifetime does not depend on a sign of the winding number $m$.
The lifetimes of \emph{wfss} with the winding number $1$ and $-1$ are same
although one couples with \emph{ns} with no singularity
and the other couples with \emph{ns} with a double quantized vortex.
The stability of \emph{wfss} is not affected by the winding number of \emph{ns}
since once an atom flips to \emph{ns} the atom is no more trapped.

When the condensate density is low,
the non-linear term $\ti{g}_1 N_0 f_w^3$ in eq.\eqref{eq-fw} is negligible.
The analytical solution is given by
\begin{align}
f_w(r)&=\frac{1}{\sqrt{\pi |m|!}}\ r^{|m|}\ee^{-r^2/2}\\
\ti{\mu}&=b^2+|m|+1\sim b^2.
\end{align}
Then $F_w(k)$ and $\tau^{-1}$ are as follows:
\begin{align}
F_w(k)&=\frac{1}{\sqrt{2\pi |m|!}}\frac{k}{b}\cdot k^{|m|}\ee^{-k^2/2},\label{Besseled_Fw}\\
\tau^{-1}&\sim 4\pi \oho \frac{(2\ti{\mu})^{|m|}}{|m|!}\ee^{-2\ti{\mu}},
\end{align}
where the approximation $\ti{\mu}\sim b^2$ is used.
The lifetime is rapidly increasing function of $\ti{\mu}$ and therefore of $b$ roughly as $\exp(2b^2)$.
When the trap geometry is $b=10$,
the condensate has such a long lifetime as $\tau\sim\ee^{200}$.
Although the singularity of the condensate shortens the lifetime,
its effect is much smaller than the exponential factor.
The condensate with a few winding number is stable after all.

When the population of trapped atoms is large,
it is necessary to solve eq.~\eqref{eq-fw} numerically.
The distribution of \emph{wfss} expand outward as $N_0$ increases.
The $b$-dependence of the lifetime is roughly same as that of analytical calculation:
the lifetime rapidly increases as $b$ increases.
To investigate its dependence upon $N_0$,
we have particularly calculated in the case of $b=7$.
Suppose $B_0=1$~G, this condition implies $B'=1.6$~T/cm and $\oho/2\pi=14$~kHz for $^{87}$Rb.
Such a large field gradient may cause
other experimental difficulty in achievement of BEC in the trap.

\begin{figure}[btp]
\begin{center}
\includegraphics[width=0.95\linewidth]{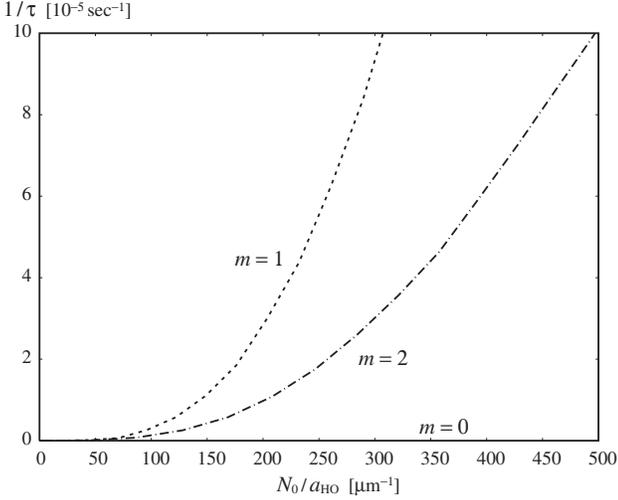}
\caption{The lifetime in trap of $b=7$;
The field is $B_0=1$~G, $B'=1.58$~T/cm and $\aho=90$~nm.
The condensate of $m=0$ is much stable and $1/\tau=0$ in this graph.
}
\label{fig-b7}\end{center}\end{figure}

Figure~\ref{fig-b7} shows our result for $1/\tau$
as a function of $N_0/\aho$, which is a number of the trapped atoms par unit length.
We have investigated condensates with $m=0, 1, 2$.
The lifetime of \emph{wfss} with $m=0$ is much longer than others
and $1/\tau\sim0$ in Fig.~\ref{fig-b7}.
The lifetime becomes shorter as $N_0$ increases in the graph.
The dependence of the lifetime upon $N_0$ is not monotonic in regions of larger $N_0$
(out of the figure).
The interaction between \emph{wfss} and \emph{ns} is affected by the distribution of $f_w(r)$,
and the condensates never decay for certain distributions.
An example for such point is 3300 atoms/$\mu$m in double quantized vortex state.

Let us compare this lifetime with the rate of three-body recombination.
In this process two spin-polarized atoms form a molecule and free from a trap.
The decay rate is given by
\begin{eqnarray}
 \frac{dn}{dt}=-Ln^3,
\label{eq-3body}
\end{eqnarray}
where $n$ is a condensate density and the recombination coefficient
for $^{87}$Rb is $L=4\times 10^{-30}$~cm$^6$/s~\cite{Moerdijk1996}.
The lifetime defined by eq.~\eqref{eq-3body} is $\tau_{\rm rec}\sim 1$~sec
at $n\sim 5\times 10^{14}$~cm$^{-3}$.
Such density corresponds to $N_0/\aho\sim50~\mu$m$^{-1}$ in Fig.~\ref{fig-b7},
where $\tau$ is much longer than 1~sec at this point.

From these considerations, as far as $b\gg 1$ the condensate is found to be stable
and the adiabatic approximation is valid.
It is also mentioned that the condensates achieved in laboratories were
considerably stable with this decay mechanism.

\subsection{Distribution over the large field gradient $b\simeq 1$}
Next we consider the case of $b\simeq 1$,
where atoms moving in a condensate feel large change of the magnetic field
and the condensate decays faster.
In this case it is impossible to use eqs.~\eqref{zeeman}, \eqref{Fw_k}.
The $b$-dependence of the lifetime is $\exp(b^2)$ approximately,
and comparable to $\tau_{\rm rec}$ at $b \sim 3$.
This is the limit that the adiabatic approximation is valid.

\begin{figure}[btp]
\begin{center}
\includegraphics[width=0.95\linewidth]{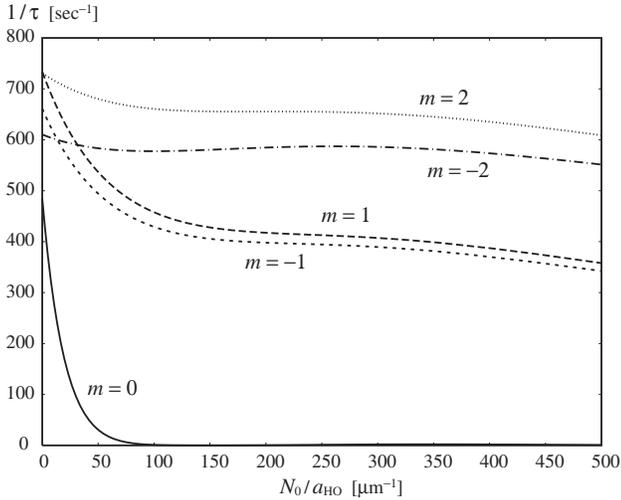}
\caption{The lifetime in trap of $b=2$;
The field is $B_0=0.1$~G, $B'=6.1$~kG/cm and $\aho=82$~nm.
The lifetime is much shorter than that in Fig.~\ref{fig-b7}.
}
\label{fig-b2}\end{center}\end{figure}

Figure~\ref{fig-b2} shows the result in a trap of $b=2$, $B_0=0.1$~G (i.e. $B'=6.1$~kG/cm) as functions of $N_0/\aho$.
The distribution size is similar as that of Fig.~\ref{fig-b7},
and $\tau_{\rm rec}\sim 1$~sec at $N_0/\aho\sim 50~\mu$m$^{-1}$.
The lifetime at this point is much smaller than 1~sec.
The decay of condensates is observable in this trap.
It should be noted in Fig.~\ref{fig-b2} that
against our intuition the lifetime becomes longer as $N_0$ increases.
The reason is that
trapping a large number of atoms brings an increase in the energy $\ti{\mu}$
as well as a change of distribution $f_w(r)$.
The $k$-dependence of $F_w(k)$ is roughly same as eq.~\eqref{Besseled_Fw}.
Hence, the increase of $\ti{\mu}$ weakens the interaction between \emph{wfss} and \emph{ns}.
The oscillation of the lifetime with a change of $N_0$ appears in Fig.~\ref{fig-b2}.

\section{Conclusion}
We have calculated the lifetime of condensate in a magnetic trap,
where field's direction spatially changes and
\emph{wfss}, \emph{ns} and \emph{sfss} interact each other.
Our investigation has confirmed that the adiabatic approximation is applicable for quantum condensates.

In the case that the condensate size is much smaller than the scale of change of field's direction $(b\gg 1)$,
the lifetime is considerably long and the condensate is stable.
The sign of $m$ does not concern the stability.
When the population of trapped atoms changes, its distribution also changes and
the lifetime oscillates as a function of $N_0/\aho$.
For certain distributions the condensates never decay with this process.

The traps used in experiments are generally $b\gg 1$ and the decay of the condensate is not observable.
The lifetime strongly depends on the energy of condensate and therefore on $b$.
It is shortened as $b$ decreases.
If one uses a trap of $b\leq 3$, the decay of the condensate will be observed.
Finally we give an example to observe this drastical change.
When one sets the field of $B_0=1$~G, $B'=6$~kG/cm $(b=11)$ and compresses 50 atoms/$\mu$m of $^{87}$Rb,
the lifetime of this condensate is determind by three-body recombination as about 1~sec.
After sweeping the bias field to 0.1~G adiabatically $(b=2)$, the lifetime becomes about 30~msec.
This decay should be observed without disturbance by recombination of atoms.
When the population of trapped atoms decreases, for instance  10 atoms/$\mu$m, the lifetime is shorten again to 3~msec.

\end{document}